\documentstyle[preprint,aps,bezier]{revtex}
\begin{document}
\draft
\preprint{
\parbox{4cm}{
\baselineskip=12pt
TMUP-HEL-9701\\ 
January, 1997\\ 
Revised 4/4/97
\hspace*{1cm}
}}
\title{A Supersymmetric Composite Model \\
       with Dynamical Supersymmetry Breaking}
\author{Noriaki Kitazawa 
\thanks{e-mail: kitazawa@phys.metro-u.ac.jp}
and Nobuchika Okada
\thanks{e-mail: n-okada@phys.metro-u.ac.jp}
\thanks{JSPS Research Fellow}}
\address{Department of Physics, Tokyo Metropolitan University,\\
         Hachioji-shi, Tokyo 192-03, Japan}
\maketitle
\begin{abstract}
We present a supersymmetric composite model with dynamical supersymmetry 
breaking.  
The model is based on the gauge group 
 $SU(2)_S \times SU(2)_H \times SU(3)_c \times SU(2)_L \times U(1)_Y$. 
Supersymmetry is dynamically broken by the non-perturbative effect of 
 the $SU(2)_S$ `supercolor' interaction. 
The large top Yukawa coupling is naturally generated 
 by the $SU(2)_H$ `hypercolor' interaction as recently proposed 
 by Nelson and Strassler. 
The supersymmetry breaking is mediated to the standard model sector 
 by a new mechanism.  
The electroweak symmetry breaking is caused by the radiative correction
 due to the large top Yukawa coupling with the supersymmetry breaking. 
This is the `radiative breaking scenario', which originates 
 from the dynamics of the supercolor and hypercolor gauge interactions. 
\end{abstract}
\newpage
%
\section{Introduction}
\label{sec:int}

In most of the supersymmetric models, if supersymmetry is broken, 
 the electroweak symmetry breaking is caused by the radiative correction 
 due to the large top Yukawa coupling. 
This mechanism is known as the `radiative breaking scenario' \cite{rbs}.   
However, to understand the scenario completely, 
 we should investigate two subjects. 
One is a mechanism of the supersymmetry breaking, 
 and the other is a natural explanation of the large top Yukawa coupling. 

The method proposed by Seiberg et al. \cite{seiberg} is remarkable, 
 when we consider the dynamical supersymmetry breaking in (N=1) 
 supersymmetric gauge theories. 
By the method, we can evaluate the non-perturbative effect 
 of a strong gauge interaction and exactly determine the dynamically 
 generated superpotential. 
There are many models [2-5] in which 
 the supersymmetry is broken by the non-perturbative effect 
 of the strong gauge interaction. 

On the other hand, the supersymmetric composite model 
 recently proposed by Nelson and Strassler \cite{ns} 
 is remarkable to understand the large top Yukawa coupling.  
They introduce the $SU(2)$ gauge interaction 
 with six doublet superfields, the `preon' superfields. 
There is $SU(6)\times U(1)_R$ global symmetry. 
The $SU(5)$ subgroup is gauged, and identified with 
 $SU(5) \supset G_{SM}$, 
 where $G_{SM}=SU(3)_c \times SU(2)_L \times U(1)_Y$ is
 the gauge group of the standard model. 
Then, the five doublet superfields transform as $\bf{5}$ of 
 the $SU(5)$, and the last one is singlet.    
Below the dynamical scale of the $SU(2)$ gauge interaction, 
 the preon superfields are confined into 
 the superfields of $\bf{5}$ and $\bf{10}$ of $SU(5)$, 
 which are identified with the up-type Higgs and the ten dimensional 
 superfields in the conventional supersymmetric $SU(5)$ GUT framework. 
The top Yukawa coupling is dynamically generated 
 by the non-perturbative effect of the $SU(2)$ gauge interaction,  
 and it can be understood as the exchange of the preon superfields 
 in microscopical point of view. 
Therefore, the top Yukawa coupling is naturally expected as $O(1)$. 

In this paper, we present a supersymmetric composite model 
 with dynamical supersymmetry breaking. 
The model is based on the gauge group 
 $SU(2)_S \times SU(2)_H \times G_{SM}$. 
We assume that dynamical scales of these gauge interaction are ordered as
 $\Lambda_S \gg \Lambda_H  \gg  E_{EW}$, where $\Lambda_S$ and $\Lambda_H$
 are the dynamical scale of the $SU(2)_S$ and $SU(2)_H$ 
 gauge interactions, respectively,
 and $E_{EW}$ is the scale of the electroweak symmetry breaking. 
It is useful to separate the model into three building blocks 
 by these energy scales, namely, 
 the dynamical supersymmetry breaking sector, 
 the preon sector proposed by Nelson and Strassler, 
 and the standard model sector. 

The dynamical supersymmetry breaking sector 
 is based on the $SU(2)_S$ `supercolor' gauge group with four doublet 
 and eight singlet superfields. 
There is $SP(2)\times U(1)_R$ global symmetry. 
The $SU(2)$ subgroup is gauged, which is identified with $SU(2)_H$.  
Therefore, there are some $SU(2)_H$ doublet superfields in this sector. 
The supersymmetry is dynamically broken at the scale $\Lambda_S$ 
 by the non-perturbative effect of the supercolor interaction. 
In addition, the $U(1)_R$ symmetry is broken by the effective K\"ahler 
 potential generated by the Yukawa coupling in the superpotential.  
Both the scalar and fermion fields in the $SU(2)_H$ doublet superfields 
 get their masses due to the breaking of the supersymmetry 
 and $U(1)_R$ symmetry.  

The preon sector is based on the $SU(2)_H$ `hypercolor' gauge group 
 with six doublet preon superfields as proposed by Nelson and Strassler. 
The large top Yukawa coupling is naturally generated 
 by the non-perturbative effect of the hypercolor gauge interaction.
Since we focus on the relation among the dynamical supersymmetry breaking, 
 the large top Yukawa coupling and the electroweak symmetry breaking, 
 we concentrate only on the third generation. 
Therefore, our model is a `toy model',  
 but it is possible to include the last two generations 
 as the elementary particles.  

The supersymmetry breaking is mediated to the preon sector 
 and the standard model sector through the $SU(2)_H \times G_{SM}$ 
 gauge interaction. 
The hypercolor gaugino and the scalar preons get the soft breaking masses 
 though radiative corrections by the $SU(2)_H$ doublet fields 
 in the dynamical supersymmetry breaking sector. 
Here, the doublet fields play a role of the `messenger fields' \cite{dn}.  
Scalar fields composed by scalar preons in the standard model 
 get masses, because of the masses of the scalar preons. 
Once the hypercolor gaugino and scalar preons become massive, 
 the standard model gauginos get masses through the radiative correction, 
 since the preon superfields have both charges 
 of the $SU(2)_H$ and $G_{SM}$.  
Here, the preon and scalar preon fields, which are the $SU(2)_H$ doublet 
 fields in the preon sector, also play a role of the `messenger fields'.   
This is a new mechanism to give the standard model gauginos 
 the soft breaking masses.  
At low energy where the preon superfields are confined, 
 the mechanism can be understood as the mixing 
 between the gauginos and the composite fermions 
 in adjoint representation. 

The electroweak symmetry is broken by the effect of the large top Yukawa 
 coupling as same as the radiative breaking scenario.  
It must be noticed that the breaking is triggered by the dynamics of 
 $SU(2)_S \times SU(2)_H$ gauge interaction, 
 since both the supersymmetry breaking and large Yukawa coupling 
 owe their origin to the dynamics. 

In section \ref{sec:cmp}, 
 we introduce the preon sector which is recently proposed 
 by Nelson and Strassler. 
In section \ref{sec:dsb},
 the dynamical supersymmetry breaking sector is constructed. 
In section \ref{sec:feed}, 
 we discuss the mediation of the supersymmetry breaking to low energy. 
In section \ref{sec:ews}, 
 the standard model sector is discussed with an emphasis on the spectrum 
 of the mediated soft breaking masses. 
In section \ref{sec:ewb}, 
 we show that the electroweak symmetry breaking really occurs in our model 
 by including the effect of the large top Yukawa coupling. 
In section \ref{sec:proton}, 
 the proton decay caused by the colored Higgs in our model is discussed, 
 and we show that there is no contradiction with the experiment. 
In section \ref{sec:axion}, 
 we briefly discuss the R-axion problem in our model 
 and give an example of the solution. 
In section \ref{sec:summary},
 we summarize our model and comment on the extension 
 of the model to include the first and second generations. 
%
%
\section{the preon sector}
\label{sec:cmp}

In this section we construct the preon sector which is based on the model 
 recently proposed by Nelson and Strassler \cite{ns}. 
Only the third generation is considered as mentioned above.  
We introduce the $SU(2)_H$ hypercolor gauge group with 
 the six doublet preon superfields. 
The maximal non-anomalous global symmetry of this sector is 
 $SU(6) \times U(1)_R$, 
 and the subgroup $SU(5) \supset G_{SM}$ is gauged, 
 where $G_{SM}=SU(3)_c \times SU(2)_L \times U(1)_Y$ 
 is nothing but the gauge group of the standard model. 
The preon superfields $P$ and $N$ transform under 
 $SU(2)_H \times SU(5)$ as follows. 
\begin{center}
\begin{tabular}{ccc}
 \hspace{1cm}& $~SU(2)_H$~ & $~SU(5)~$   \\
$ P $        &  \bf{2}     &  \bf{5}     \\
$ N $        &  \bf{2}     &  \bf{1}    
\end{tabular}
\end{center}
It is convenient to decompose the preon superfield $P$ as $P=(d\; h)$,  
 where $d$ and $h$ transform as
 $({\bf{3}},{\bf{1}},-1/3)$ and $({\bf{1}},{\bf{2}},1/2)$, respectively, 
 under the standard model gauge group $G_{SM}$.   

In addition, we introduce two elementary superfields 
 $\Phi_1$ and $\Phi_2$, both of which are in $\bf{5}^*$ representation 
 of the $SU(5)$. 
The field $\Phi_1=(\overline{b} \; \overline{\ell})$ represents 
 the superfields of the anti-bottom quark 
 and the lepton doublet $\overline{\ell}=(\tau\; -\nu_{\tau})$, and 
 $\Phi_2$ is corresponds to the down-type Higgs superfield  
 $\Phi_2=(\overline{D} \; \overline{H})$ with 
 $\overline{D}=({\bf{3}}^{*},{\bf{1}},1/3)$ and 
 $\overline{H}=({\bf{1}},{\bf{2}},-1/2)$.  

At high energy where the hypercolor interaction is weak, 
the superpotential is generally given by (including dimension 4 operators)
\begin{eqnarray} 
W &= &\eta {\overline{H}}[hN]+ \eta_D {\overline{D}}[dN] 
\label{nstree} \nonumber \\
 &+&\frac{\kappa_b}{M_b}{\overline{H}}[dh]{\overline{b}} 
  +\frac{\kappa_{\ell}}{M_{\ell}}{\overline{H}}[hh]\ell \; ,
\end{eqnarray} 
where square brackets denote the contraction of the $SU(2)_H$ index  
 by the $\epsilon$-tensor, 
 and $M_b$ and $M_{\ell}$ are parameters with dimension one.  
Although the third and fourth terms in eq.(\ref{nstree}) are needed 
 to generate the bottom and tau Yukawa couplings,  
 we neglect these terms in the following, and consider only 
 the top Yukawa coupling.   

Below the scale $\Lambda_H$ where the $SU(2)_H$ interaction 
 becomes strong, this theory is described by the massless 
 $SU(2)_H$ singlet effective superfields $M_{ij}$, 
 where $i$ and $j$ are flavor indices.   
The effective superfield is the 15-component antisymmetric tensor 
 given by 
\begin{equation}
M=\epsilon^{\alpha \beta}
\left( \begin{array}{c}
d\\
h\\
N\\
\end{array}\right)_{\alpha}
{\left( 
\begin{array}{ccc}
d & h &N 
\end{array}\right)_{\beta}}
=\left( \begin{array}{ccc}
[dd] & [dh] & [dN]   \\ 
$[$hd] & [hh] & [hN] \\
$[$Nd] & [Nh] &  0    
\end{array}
\right) \; ,  
\end{equation}
where $\alpha$ and $\beta$ are the $SU(2)_H$ indices.  
This tensor is decomposed as $\bf{5}+\bf{10}$ of $SU(5)$.  
\begin{equation}
\bf{5}:
\left(\begin{array}{c}
  [dN] \\
$[$hN]
\end{array}
\right)
\; , \; \; \; \;  
\bf{10}:
\left( \begin{array}{cc}
  [dd] & [dh]   \\
$[$hd] & [hh]
\end{array}
\right) \; .
\end{equation}
Here, the superfield in $\bf{5}$ representation 
 can be identified with the up-type Higgs superfield 
 $([dN]\; [hN]) \sim \Lambda_H (D\; H)$, 
 where $D=({\bf{3}},{\bf{1}},-1/3)$ and $H=({\bf{1}},{\bf{2}},1/2)$. 
The components of the superfield in $\bf{10}$ representation 
 are also identified as 
 $[dd]\sim \Lambda_H \overline{t}$, $[dh] \sim 
\Lambda_H (t\; b)=\Lambda_H q$, 
 and $[hh]\sim \Lambda_H \overline{\tau}$, respectively. 

The dynamically generated superpotential is exactly given by 
\begin{equation}
W_{dyn}=-\frac{1}{\Lambda^3_H}$Pf$(M)
=-\displaystyle{\frac{1}{\Lambda^3_H} 
\frac{1}{2^3 3!}} \epsilon^{ijklmn}
M_{ij}M_{kl}M_{mn} \; .  \label{nsdy}
\end{equation}
>From eqs.(\ref{nstree}) and (\ref{nsdy}), 
 the total effective superpotential is given by 
\begin{eqnarray}
W_{eff}
&=&\alpha Hq\overline{t} + \beta qqD +\gamma \overline{t}D \overline{\tau}
  \label{nseff} \nonumber \\
&+& \mu \overline{H}H +\mu_D \overline{D} D \; .
\end{eqnarray}
Note that $\alpha \sim \beta \sim \gamma \sim O(1)$ is naturally expected, 
 since it can be understood that 
 these Yukawa couplings are constructed 
 by the exchange of the preon superfields. 
The first term in eq.(\ref{nseff}) is nothing but the top Yukawa coupling, 
 which is naturally large. 
The fourth and fifth terms are the $\mu$-terms of the $SU(2)_L$ doublet 
 and the color triplet Higgs superfields, respectively,  
 where $\mu \sim \Lambda_H \eta$, and $\mu_D  \sim \Lambda_H \eta_D$ 
 from eq.(\ref{nstree}).  

Note that this composite model is the only one 
 which satisfy the following three requirements. 
The first is that the model is based on the gauge group $SU(N)$ with vector-like $N_f$ 
 matter fields in the fundamental representation.  
The second is that there is only the `mesonic' effective superfields 
 but not the `baryonic' one. 
The third is that the Yukawa coupling has to be dynamically generated.   
These can be satisfied only if $N_f=N+1$ and $N=2$. 
%
%
\section{the Dynamical supersymmetry Breaking sector}
\label{sec:dsb}

In order for the supersymmetric composite model discussed 
 in the previous section 
 to be realistic, the supersymmetry should be broken. 
We construct the dynamical supersymmetry breaking sector in this section.

It is recently pointed out that the theory of 
 the $SU(2)$ gauge interaction with four doublet and six singlet  
 superfields breaks supersymmetry by the non-perturbative effect of 
 the gauge interaction \cite{iy}.  
The dynamical supersymmetry breaking sector 
 in our model has the same structure as the theory. 

This sector is based on the $SU(2)_S$ supercolor gauge group 
 with four doublet and eight singlet superfields. 
The maximal non-anomalous global symmetry of this sector   
 is $SP(2) \times U(1)_R$.  
The $SU(2)$ subgroup of $SP(2)$ is gauged, and identified 
 with the hypercolor gauge group $SU(2)_H$.  
Note that we treat the hypercolor interaction as  weak 
 in the discussion of the dynamical symmetry breaking, 
 because of the assumption $\Lambda_S \gg \Lambda_H$. 

The particle contents are as follows. 
\begin{center}
 \begin{tabular}{ccccc}
  \hspace{1cm} & $ ~SU(2)_S~$ & $ ~SU(2)_H~ $ & $ ~U(1)_R~ $ \\
$ Q $           &   \bf{2}     &   \bf{2}      &      0       \\
$ \tilde{Q}_1 $ &   \bf{2}     &   \bf{1}      &      0       \\ 
$ \tilde{Q}_2 $ &   \bf{2}     &   \bf{1}      &      0       \\
$ Z $           &   \bf{1}     &   \bf{1}      &      2       \\
$ Z^{'}$        &   \bf{1}     &   \bf{1}      &      2       \\
$ Z_1 $         &   \bf{1}     &   \bf{2}      &      2       \\
$ Z_2 $         &   \bf{1}     &   \bf{2}      &      2       \\
$ X $           &   \bf{1}     &   \bf{1}      &      0       \\
$ Y $           &   \bf{1}     &   \bf{1}      &      2     
 \end{tabular}
\end{center}
The tree level superpotential is introduced as 
\begin{equation}
W_{tree}=\lambda \; $tr$[\hat{Z} \hat{V}] + 4 \lambda_V Z V 
          +\lambda_{Z} Z X^2 +\lambda_Y Y X^2\;,  \label{tree}
\end{equation}
where $\hat{Z}$ and $\hat{V}$ are the antisymmetric $4 \times 4$ matrices 
 defined by 
\begin{equation}
\hat{Z}=
\left[ \begin{array}{cccc}
  0     &  ~~Z+Z^{'} ~~  & {Z_1}^1            &    {Z_2}^1       \\ 
~~-(Z+Z^{'})~~  & 0          & {Z_1}^2            &    {Z_2}^2       \\
 -{Z_1}^1      & -{Z_1}^2       &  0                &  ~~ Z-Z^{'} ~~  \\
 -{Z_2}^1      & -{Z_2}^2       &   ~~ -(Z-Z^{'})~~ &  0              \\ 
\end{array}
\right]
\end{equation}
and 
\begin{eqnarray}
\hat{V}&=&\epsilon^{\alpha \beta}
\left( \begin{array}{c}
Q\\
\tilde{Q}_1\\
\tilde{Q}_2\\
\end{array}\right)_{\alpha}
{\left( 
\begin{array}{ccc}
Q & \tilde{Q}_1 & \tilde{Q}_2  
\end{array}\right)_{\beta}}    \nonumber  \\
&=&
\left[ \begin{array}{cccc}
  0            &  ~~V+V^{'} ~~  &   V_{11}          &    V_{21}     \\
~~-(V+V^{'})~~   &     0        &   V_{12}          &    V_{22}     \\
 -V_{11}        & -V_{12}        &     0             & ~~V-V^{'} ~~  \\
 -V_{21}          & -V_{22}        &   ~~ -(V-V^{'})~~ & 0             
\end{array}
\right] \;  ,
\end{eqnarray}
 respectively. 
Here, $a=1,2$ of ${Z_i}^a$ and $V_{i a}$ ($i=1,2$) is 
 the $SU(2)_H$ index, and 
 $\alpha$ and $\beta$ are the $SU(2)_S$ indices. 
The field $\hat{Z}$ is in the reducible representation ${\bf{1}}+{\bf{5}}$ 
 of $SP(2)$:  
$Z$ is singlet, and $Z^{'}$ and $Z_i$ are in $\bf{5}$. 
This is the same for the field $\hat{V}$. 
Note that eq.(\ref{tree}) is the general superpotential 
 which possesses the $SP(2)\times U(1)_R$ global symmetry. 

Below the scale $\Lambda_S$ where the supercolor interaction 
 becomes strong, 
 this sector is described by the effective superfield $\hat{V}$.  
The dynamically generated superpotential is exactly given by 
\begin{eqnarray}
W_{dyn}
&=& 
A ( \mbox{Pf} \hat{V}-{\Lambda_S}^4)    \label{dyn}  \nonumber \\
&=&
A (V^2-V^{'2}- [ V_1 \; V_2 ]-{\Lambda_S}^4 ) \; ,
\end{eqnarray}
where $A$ is the Lagrange multiplier superfield, 
 and the square brackets denote 
 the contraction of the $SU(2)_H$ indices with the $\epsilon$-tensor. 
The total effective superpotential is given by 
 $W_{eff}=W_{tree}+W_{dyn}$.  

In the analysis of the vacuum, eq.(\ref{dyn}) is regarded as a constraint, 
 Pf$\hat{V}-{\Lambda_S}^4=0$.  
By this constraint one superfield should be eliminated from $W_{eff}$.  
We eliminate the $SP(2)$ singlet superfield $V$ from $W_{eff}$ 
 by the constraint, 
 since it is expected that the vacuum is realized 
 with the largest global symmetry.  
Then, the effective superpotential is rewritten by 
\begin{eqnarray}
W_{eff}= &-& 4 (\lambda - \lambda_V )\; Z \; 
 \sqrt{{\Lambda_S}^4+V^{'2}+ [V_1 \;V_2] }  
 \label{eff1} \nonumber   \\ 
&-&\lambda  \left \{ 4 Z^{'} V^{'} +2  [Z_1 \; V_1] 
 + 2 [Z_2 \; V_2] \right \}
+\lambda_Z Z X^2 +\lambda_Y Y X^2  \; , 
\end{eqnarray}
where the ambiguity of the sign in front of the square root 
 is absorbed by the $U(1)_R$ phase rotation. 
By factoring out $\Lambda_S$ to normalize the dimension 
 of the effective superfields, 
 and expanding the square root, we obtain 
\begin{eqnarray} 
W_{eff} \simeq &-&4 (\lambda -\lambda_V) \; Z \; \left\{  
{\Lambda_S}^2+ \frac{1}{2} V^{'2}+ \frac{1}{2} [V_1 \; V_2] 
\right\}    \label{eff2} \nonumber  \\
&-& \lambda \Lambda_S \left\{
 4 Z^{'} V^{'} +2 [Z_1 \; V_1] +2 [Z_2 \; V_2] \right\} 
+ \lambda_Z  Z X^2 +\lambda_Y Y X^2  \; .
\end{eqnarray}
This effective superpotential is one of the type 
 of O'Raifeartaigh models \cite{o'r}, 
 and five supersymmetric vacuum conditions,  
 $\partial W_{eff}/\partial Z=0$, $\partial W_{eff}/\partial Z^{'}=0 $, 
 $\partial W_{eff}/\partial Z_i=0 \; (i=1,2)$ 
 and $\partial W_{eff}/\partial Y=0 $, 
 cannot be satisfied simultaneously. 
As a result, the supersymmetry is broken. 

Note that the effective potential has the `pseudo-flat direction' 
 as O'Raifeartaigh models have. 
For example, the vacuum energy remains minimum along the direction 
 of arbitrary vacuum expectation value $\langle Z \rangle$ 
\footnote{
We use the same notation for the superfield itself 
 and the scalar component of the superfield.  
}
, when the other fields have no vacuum expectation values. 
However, this flatness will disappear, if the quantum correction 
 to the K\"ahler potential is considered. 
The effective K\"ahler potential at low energy 
 cannot be reliably evaluated, 
 because of the strong supercolor interaction.   
Therefore, at the high energy where the supercolor is weak and negligible, 
 we evaluate it by considering only the Yukawa couplings 
 in the tree level superpotential.
Note that the superfield $Z$ couples to the massless superfield $X$   
 in eq.(\ref{tree}). 

In the non-supersymmetric massless $\lambda \phi^4$ theory, 
 it is well known that the infrared divergence appears in 
 the loop correction of the massless field. 
Although the infrared divergence disappears 
 with summing up the 1-loop diagrams, 
 the potential with the 1-loop correction becomes singular at the point 
 $\langle \phi \rangle =0$, as a result. 
The problem of the infrared divergence or 
 the singularity at the origin is solved 
 by the fact that the non-zero vacuum expectation value 
 $\langle \phi \rangle \neq 0$
 emerges in the effective potential. 
This mechanism is pointed out by Coleman and Weinberg \cite{cw}.

When the effective K\"ahler potential of the superfield $Z$ is evaluated 
 by considering the loop correction of the massless superfield $X$, 
 we encounter the same situation with the infrared divergence, 
 since the superfield is treated as the boson field 
 in the supergraph method \cite{sgraph}. 
Unfortunately, the 1-loop effective K\"ahler potential 
 is not reliable (see Appendix).  
However,  we can expect that the effective K\"ahler potential of $Z$ 
 is non-trivial, and the minimum of the effective potential 
 is realized at the point $\langle Z \rangle \neq 0$ 
\footnote{
The possibility that the scalar component of $Z$ has 
 a large vacuum expectation value 
 without the additional Yukawa coupling $\lambda_Z$ 
 is pointed out by Shirman \cite{shirman}.  
If it is true, we can start a more simple $SP(2) \times U(1)_R$ 
 symmetric superpotential, 
 $W_{tree}= \lambda \;  \mbox{tr} [\hat{Z} \hat{V}] + 4 \lambda_V Z V$,  
 without introducing the superfields $X$ and $Y$. 
}
, according to the mechanism of Coleman and Weinberg. 
This discussion is also applied to the effective K\"ahler potential 
 of the superfield $Y$, 
 and $\langle Y \rangle \neq 0$ is also expected. 

In the following, we assume that K\"ahler potentials 
 of all the fields except for 
 $Z$ and $Y$ are naive,  
 and their scalar components have no vacuum expectation values.   
By this assumption, the vacuum energy is given by 
\begin{equation}
E_{vac}= 16 (\lambda -\lambda_V)^2 \Lambda_S^4 \Bigg/ \left[ 
\frac{\partial}{\partial (Z^{\dagger} Z)} 
\left( (Z^{\dagger} Z ) 
\frac{\partial K_Z}{\partial (Z^{\dagger} Z)} 
 \right)  \right]_{Z=\langle Z \rangle}  \; ,
\end{equation}
where $K_Z$ is the non-trivial K\"ahler potential of the superfield $Z$. 
The supersymmetry is broken by the $SU(2)_S$ supercolor gauge dynamics, 
 unless the denominator becomes singular. 
%
%
\section{The Mediation of Supersymmetry breaking}
\label{sec:feed}

In this section, we investigate the mediation 
 of the supersymmetry breaking. 
The breaking is mediated to the preon sector and the standard model sector 
 by the radiative corrections due to 
 the $SU(2)_H$ and $G_{SM}$ charged particles. 
The $SU(2)_H$ doublet fields play the role 
 of the `messenger fields' \cite{dn}. 

Let us consider the mass spectrum of the $SU(2)_H$ doublet fields 
 from the superpotential of eq.(\ref{eff2}). 
These are given by 
\begin{eqnarray}  
{\cal L}_{mass} &=& 
-(m^2+m^{'2})(V_1^{\dagger} V_1+V_2^{\dagger} V_2) 
-F\; [V_1 \; V_2] +h.c.  \label{mass} \\  \nonumber 
&-& m^{'2} (Z_1^{\dagger} Z_1 +Z_2^{\dagger} Z_2) 
-m m^{'} (Z_1^{\dagger} V_2 +Z_2^{\dagger} V_1) +h.c. \\ \nonumber 
&-&m\; [\psi_{V_1} \; \psi_{V_2}] 
-m^{'}([\psi_{Z_1} \; \psi_{V_1}]+ [\psi_{Z_2} \; \psi_{V_2}] )+ h.c. \; , 
\end{eqnarray}
where 
\begin{eqnarray}
F &\equiv& 8(\lambda -\lambda_V)^2 \Lambda_S^2 \Bigg/ \left[
\frac{\partial}{\partial (Z^{\dagger} Z)} 
\left( (Z^{\dagger} Z ) 
\frac{\partial K_Z}{\partial (Z^{\dagger} Z)} 
 \right) \right]_{Z=\langle Z \rangle} \; ,\\ \nonumber 
m &\equiv& 2 (\lambda -\lambda_V) \langle Z \rangle \; , \\ \nonumber 
m^{'} &\equiv& 2\lambda \Lambda_S\; .  
\end{eqnarray}
We can regard $F$, $m$ and $m^{'}$ as three independent free parameters 
 instead of the four independent free parameters  
 $\lambda$, $\lambda_V$, $\lambda_Z$ and $\Lambda_S$ in the following
\footnote{
Note that the K\"ahler potential $K_Z$ and the vacuum expectation value 
 $\langle Z \rangle$ are determined by the coupling $\lambda_Z$. 
}
.  
Note that the $SU(2)_H$ doublet superfields $Z_i$ and $V_i$ ($i=1,2$) 
 should be decoupled at the energy scale of $\Lambda_H$,   
 in order for the mechanism by Nelson and Strassler really to works. 
This fact demands to take $\Lambda_H^2 \ll  F, m^2$, and $m^{'2}$. 
In the following discussion, we set $m^{'} \ll m$ and $F \ll m^2$, 
 and treat the $SU(2)_H$ coupling perturbativly. 

The hypercolor gaugino gets its soft breaking mass 
 through the 1-loop diagram in Fig.1. 
It is given by 
\begin{equation}
m_{\lambda_H} \sim \frac{\alpha_{H}}{8 \pi} \frac{F}{m}\; , \label{hgmass}
\end{equation}
where $\alpha_{H}=g_H^2/4\pi$ is the coupling constant 
 of the hypercolor interaction.  
This is the first order correction of $\alpha_{H}$. 

The scalar preons, $\tilde{P}$ and $\tilde{N}$, also 
 get their masses through the 2-loop diagrams in Fig.2.  
Since their $SU(2)_H$ charge is the same, 
 they have the same masses given by 
\begin{equation}
m_{\tilde{P}}^2=m_{\tilde{N}}^2 \sim \frac{3}{2} 
\left( \frac{\alpha_H}{4 \pi} \right)^2 \left( \frac{F}{m} \right)^2 
\label{spmass} \; .  
\end{equation}
This is the second order correction of $\alpha_H$. 
Note that the scalar preon masses are the same order 
 of the hypercolor gaugino mass.   
Since the scalar preons have the masses, 
 the composite scalars, the scalar quarks 
 $\tilde{q}$ and $\tilde{\overline{t}}$, 
 the scalar lepton $\tilde{\overline{\tau}}$, 
 and the up-type Higgs boson also have the same soft breaking masses. 
We assume that their masses of the composite scalars are twice of 
 the scalar preon mass, for simplicity.  

Furthermore, the gauginos in the standard model get their masses through 
 the radiative correction including the preon superfield $P$, 
 since it has not only $SU(2)_H$ charge but also $G_{SM}$ charge. 
By 3-loop diagram 
\footnote{ 
 A similar diagram was considered by L. Randall in ref.\cite{randall}. 
}
in Fig.3, the mass is given by 
\begin{equation}
m_{\lambda_N} \sim \frac{3 \alpha_N}{8 \pi} 
 \left( \frac{\alpha_H}{4 \pi} \right)^2 
\frac{F}{m} \; \ln\left(\frac{m}{\Lambda_H}
\right) \label{ewg} \; ,
\end{equation}
where $\alpha_N$ ($N=1,2,3$) is the coupling constant 
 of the $SU(N)$ interaction in the standard model  
\footnote{
In this calculation, the infrared divergence occurs, 
 which means that the mass at zero momentum cannot be defined. 
We have taken $\Lambda_H$ as the cut off parameter 
 of the infrared divergence,  
 since the preon superfield $P$ are confined at the energy lower 
 than $\Lambda_H$. 
}
. 
This 3-loop correction is the second order for $\alpha_H$, 
 and the first for $\alpha_N$.   

Note that the preon and scalar preon fields play 
 the role of `messenger fields', 
 and this is a new mechanism to give soft breaking masses 
 to the standard model gauginos. 
At low energy where the preon and scalar preon are confined, 
 the mechanism can be understood as the mixing between gauginos 
 and the composite fermions in adjoint representation.  
This is a special picture of this supersymmetric composite model. 

We have shown that the soft breaking masses of the particles 
 in the standard model are really generated. 
However, in the discussion of the electroweak symmetry breaking, 
 the above perturbative estimation cannot be reliable, 
 since the hypercolor interaction becomes strong at low energy. 
In the following, we use the masses given in eqs.(\ref{spmass}) 
 and (\ref{ewg}) with $\alpha_H/4 \pi =1$ for order estimation.  
The condition $m_{\lambda_H}, m_{\tilde{P}} \ll \Lambda_H$ must 
 be satisfied in order for the mechanism by Nelson and Strassler 
 really to work at low energy.
%
%
\section{The Standard Model Sector}
\label{sec:ews}

Let us consider the standard model sector at the scale 
 where the preon superfields are confined by the $SU(2)_H$ interaction. 
The large top Yukawa coupling is generated 
 as discussed in section \ref{sec:cmp}. 
The scalar quarks ($\tilde{q}$ and $\tilde{\overline{t}}$), 
 the scalar lepton ($\tilde{\overline{\tau}}$) 
 and the up-type Higgs boson have the same soft breaking masses 
\begin{equation}  
 m_{\tilde{q}}=m_{\tilde{\overline{t}}}=
 m_{\tilde{\overline{\tau}}}= m_{H} \sim 2 m_{\tilde{P}} 
 \sim  \sqrt{6} \left( \frac{F}{m} \right)\; ,  \label{sqmass}
\end{equation}
where we use eq.(\ref{spmass}) with $\alpha_H/4 \pi =1$.  
The standard model gauginos have masses as given by eq.(\ref{ewg}) 
 with $\alpha_H/4 \pi =1$.  
Note that the ratio of the masses of two gauginos are determined by 
 the ratio of two gauge coupling; 
\begin{equation}
m_{\lambda_{N}}/ m_{\lambda_{M}} = 
 \alpha_N /   \alpha_M  \; , \label{relation}
\end{equation}
where $N$,$M = 1,2,3$.   
This relation is the same as obtained in the gauge-mediate model 
 of ref.\cite{dn}.

Furthermore, other soft breaking masses in the standard model 
 are also induced by the radiative correction through the standard model 
 gauge interaction. 
By the 1-loop diagram in Fig.4, the mass of the down-type Higgs boson 
 $\overline{H}$ is given by 
\begin{equation}
m_{\overline{H}}^2 
 \sim \frac{\alpha_1}{8 \pi} \left[ 
   \mu_D^2 \; \ln  \left( 1+\frac{m_{\tilde{q}}^2}{\mu_D^2}  \right) 
+  m_{\tilde{q}}^2  \; \ln \left(
\frac{\mu_D^2+m_{\tilde{q}}^2}{\mu^2+m_{\tilde{q}}^2} \right) 
- \mu^2 \; \ln \left( 1+\frac{m_{\tilde{q}}^2}{\mu^2}  \right) \right]
\label{hbarsoft}  \; .   
\end{equation}
The soft breaking term $B\overline{H} H + h.c.$ 
 in the Higgs potential is also induced by 
 the 1-loop diagram in Fig.5.
We obtain 
\begin{equation}
B \sim \frac{3 \alpha_2}{2 \pi} m_{\lambda_2} \; \mu \ln \left( 
\frac{\Lambda_H}{m_{\lambda_2}}
 \right)  \label{bterm} \; , 
\end{equation}
where the ultraviolet divergence is cut off 
 by the composite scale $\Lambda_H$, 
 and we assume $\mu^2 / m_{\lambda_2}^2 \ll 1$. 

By considering all the estimated soft breaking masses, 
 the Higgs potential is given by 
\begin{equation}
V= \frac{1}{2} (\mu^2+m^2_{\overline{H}}) \overline{h}^2 
   + \frac{1}{2} (\mu^2+m^2_{\tilde{q}}) h^2  
  + B \overline{h} h + \frac{g_1^2 + g_2^2}{32} (\overline{h}^2- h^2)^2 
\; , \label{ewtree}
\end{equation}
where $g_1$ and $g_2$ are the gauge coupling constants, 
 and $\overline{h}$ and $h$ 
 are the real part of the neutral component 
 of $\overline{H}$ and $H$, respectively. 
Since $m^2_{\overline{H}} > 0$ and $B < m^2_{\tilde{q}}$ 
 from eqs.(\ref{ewg}), (\ref{sqmass}), (\ref{hbarsoft}), and (\ref{bterm}) 
 with $\mu^2 < \mu_D^2$, 
 the condition  $B^2  < (\mu^2+m^2_{\overline{H}})(\mu^2+m^2_{\tilde{q}})$ 
 is satisfied. 
Therefore, the electroweak symmetry cannot be broken at this level. 
%
\section{The Electroweak Symmetry Breaking }
\label{sec:ewb}

We have seen that the electroweak symmetry cannot be 
 broken in eq.(\ref{ewtree}). 
However, if we consider the radiative corrections due to 
 the large top Yukawa coupling, 
 the electroweak symmetry breaking really emerges 
 according to the `radiative breaking scenario' \cite{rbs}. 

Let us consider the 1-loop radiative corrections induced 
 by the top and scalar top loops. 
The 1-loop effective potential is given by 
\begin{equation}
V_{1-loop}= \frac{3}{8 \pi^2} \int^{\Lambda_H^2}_{0} dk^2
k^2 \left[ 
\ln 
\left( 1+ \frac{g_t^2 h^2 /2}{k^2 +m_{\tilde{q}}^2} \right)
- \ln  \left( 1+ \frac{g_t^2 h^2 /2}{k^2}  \right)
 \right] \; , \label{1loop}
\end{equation}
where we use $\Lambda_H$ as a cut off parameter 
 because of the compositeness of the 
 top, the scalar top and the up-type Higgs doublet. 
In eq.(\ref{1loop}), the first and second terms 
 in the square brackets are given by 
 the loop corrections of the scalar top and top quark, respectively. 
One can check $V_{1-loop} \rightarrow 0$ in the supersymmetric limit 
 $m_{\tilde{q}}^2 \rightarrow 0$. 
Now, we obtain the effective potential $V_{eff}=V +V_{1-loop}$ 
 from eq.(\ref{ewtree}) and (\ref{1loop}). 

In order for the electroweak symmetry to be broken, 
 two minimization conditions, $\partial V_{eff}/ \partial \overline{h}=0$ 
 and $\partial V_{eff}/ \partial h =0$, should be satisfied 
 with non-zero values of $\langle \overline{h}\rangle$ 
 and $\langle h \rangle$. 
Furthermore, these vacuum expectation values should satisfy the condition, 
 $\langle \overline{h} \rangle^2 + \langle h \rangle^2 = v^2$, 
 to realize the correct scale of the electroweak symmetry breaking, 
 where $v \simeq 246$ GeV.  
By considering this condition, two minimization conditions 
 are described by 
\begin{equation}
\frac{\partial V_{eff}}{ \partial \overline{h}} =0 
\; \Rightarrow \; 
\mu^2 + m_{\overline{H}}^2 + \frac{1}{2} M_Z^2 \cos 2\beta  
+ B \tan \beta  = 0 \; , \label{v1}
\end{equation}
and 
\begin{eqnarray}
\frac{\partial V_{eff}}{ \partial h} = 0 
\;& \Rightarrow & \;  
\mu^2  +m_{\tilde{q}}^2  - \frac{1}{2} M_Z^2 \cos 2\beta +B \cot \beta
 \label{v2}   \nonumber \\
&+& \frac{3}{4 \pi^2} \frac{m_t^2}{v^2 \sin^2 \beta} \left[ 
  m_t^2           \ln \left(\frac{m_{\tilde{q}}^2 + m_t^2 }{m_t^2}\right) 
- m_{\tilde{q}}^2 \ln \left(\frac{\Lambda_H^2}{m_{\tilde{q}}^2 + m_t^2} \right)
\right]= 0  \; ,
\end{eqnarray}
where we use the definitions, 
 $\langle \overline{h} \rangle = v \cos \beta$, 
 $\langle h \rangle = v \sin \beta$, 
 $m_t^2=g_t^2 \langle h \rangle^2 /2$ for top quark mass, and 
 $M_Z^2=(g_1^2+g_2^2)v^2/4$ for $Z$ boson mass. 
Note that $\mu < 0 $ is required for $\tan \beta$ to be positive 
 from eqs.(\ref{bterm}) and (\ref{v1}). 

Now we show that
 there is a realistic solution by substituting the concrete values
 to the parameters in eqs.(\ref{v1}) and (\ref{v2}). 
There are five independent unknown parameters in the two equations: 
 $\mu$, $m_{\bar H}$, $m_{\tilde q}$, $B$ and $\tan \beta$.
These parameters are described by the six parameters,
 $\mu$, $\mu_D$, $m_{\tilde q}$, $\Lambda_H$, $r \equiv m/\Lambda_H$
 and $\tan \beta$ through 
 the eqs.(\ref{ewg}), (\ref{sqmass}), (\ref{hbarsoft}), and (\ref{bterm}). 
We set the values of the four parameters as
 $\mu = -200$GeV, $\mu_D = 1$TeV, $m_{\tilde q} = 20$TeV, and $r = 1000$.
The values of $\mu_D$ and $m_{\tilde q}$ have to be large enough,
 since there is no evidence of the production 
 of the exotic colored particle.
The value of $r$ has been set as $r \gg 1$ to keep the hierarchy
 which is assumed in the previous section.
The remaining two parameters, $\Lambda_H$ and $\tan \beta$,
 are determined by solving the eqs.(\ref{v1}) and (\ref{v2}).
By numerical calculation, we can find the solution
\begin{equation}
\tan \beta \sim 3.7 \; , \; \; 
\Lambda_H/m_{\tilde{q}} \sim 9.3 \times 10^4 
\end{equation}
with $\alpha_3=0.12$, $\alpha_2=0.033$, $\alpha_1=0.017$, 
 $m_t=180$ GeV and $v=246$ GeV. 
Note that the hierarchy $m_{\lambda_H}, m_{\tilde{P}} \ll \Lambda_H$, 
 which must be satisfied to work the mechanism proposed 
 by Nelson and Strassler, is realized, because of 
 $m_{\lambda_H} \sim m_{\tilde{P}} \sim m_{\tilde{q}} \ll \Lambda_H$ 
 (see eqs.(\ref{hgmass}), (\ref{spmass}) and (\ref{sqmass}) ). 
As a result, the compositeness scale is very high.  

The values of $\tan \beta$ and $\Lambda_H / m_{\tilde{q}}$ 
 are the increasing functions
 of $|\mu |$ and $\mu_D$, while the decreasing functions 
 of $m_{\tilde{q}}$ and $r$. 
That parameter dependence of the solution can be seen analytically 
 by the rough estimation of the solution. 
We regard eq.(\ref{v1}) as the equation to determine the value 
 of $\tan \beta$. 
>From eqs.(\ref{ewg}), (\ref{sqmass}) and (\ref{bterm}), we obtain 
\begin{equation}
\tan \beta \sim -(\mu^2+ m_{\overline{H}}^2) /B
\sim - (\mu^2+ m_{\overline{H}}^2) \Bigg/ 
\left[ 
\frac{3\sqrt{6} \alpha_2^2}{32 \pi^2} m_{\tilde{q}} \; \mu \;  \ln(r)
\right]   \label{tan} \; ,
\end{equation}
where the term $1/2 M_Z^2 \cos 2\beta$ is neglected with $\mu=-200$ GeV, 
 and the logarithm in eq.(\ref{bterm}) is considered as $O(1)$. 
Substituting eq.(\ref{hbarsoft}) to (\ref{tan}), 
 one can check the parameter dependence of $\tan \beta$ mentioned above.  
On the other hand, we obtain from eq.(\ref{v2}) 
\begin{equation}
\frac{\Lambda_H}{m_{\tilde{q}}} \sim
\exp \left[ 
\frac{2 \pi^2 v^2}{3 m_t^2}
\frac{\tan^2 \beta}{1+\tan^2 \beta} 
\right] \; , \label{ratio} 
\end{equation}
where the relation $m^2_{\tilde{q}} \gg \mu^2$, $M^2_Z$, and $B$  is used. 
The value of $\Lambda_H /m_{\tilde{q}}$ has the same dependence 
 as $\tan \beta$, 
 since it is the increasing function of $\tan \beta$.  

By using $m_{\tilde{q}}$, the mass of the gauginos 
 in the standard model sector are given by 
\begin{equation} 
  m_{\lambda_N} \sim \frac{\sqrt{6} \alpha_N}{ {16 \pi}} m_{\tilde{q}}
  \; \ln(r) \;. 
\end{equation}
Then, we obtain the values of the masses 
 of the gauginos for $m_{\tilde{q}}=20$ TeV as 
\begin{equation}
m_{\lambda_3}\sim 810 \mbox{GeV}\; , \;  
 m_{\lambda_2} \sim 220 \mbox{GeV} \; , \; 
m_{\lambda_1} \sim 110 \mbox{GeV} \; .
\end{equation}
These values are experimentally acceptable \cite{pdg}. 
%
%
\section{The proton decay problem}
\label{sec:proton}
In our model there are colored Higgs superfields $\overline{D}$ and $D$ 
 which may cause the proton decay. 
The baryon number violating interaction at the tree level 
 in the first and second generations is forbidden 
 by imposing usual $U(1)_{B-L}$ symmetry, 
 under which $\overline{D}$ and $D$ are singlets. 
On the other hand, there is another $U(1)_{B-L}$ symmetry 
 in the third generation, 
 under which $\overline{D}$ and $D$ have charge $2/3$ and $-2/3$, 
 respectively. 
This symmetry allows the baryon number violating interactions 
 as the second and third terms in eq.(\ref{nseff}). 
These interactions may cause too rapid proton decay 
 through the flavor mixing between the third and  other generations 
 which reduces these two symmetries to the unique $U(1)_{B-L}$ symmetry
\footnote{ It is possible to identify $\overline{\tau}$ 
 in eq.(\ref{nseff}) with a singlet heavy particle instead 
 of the right-handed tau, as done in the original paper 
 by Nelson and Strassler \cite{ns}. 
 In this case, there in no problem on the proton decay. 
}. 

To discuss the problem of proton decay, 
 we should know the quark and lepton mass matrices, in other words, 
 we need a model which explains the origin 
 of masses of all quarks and leptons. 
Although such a model is not yet constructed, 
 we can discuss proton decay with some assumptions 
 which can be reasonably expected in our model. 

In order to generate the up quark mass and the mixing 
 between the up and top quarks, the following non-renormalizable 
 interactions must be introduced in the superpotential.  
\begin{eqnarray}
W &=& \frac{\kappa_{uu}}{M_{uu}} [hN] q_1 \overline{u} 
 +\frac{\kappa_{ut}}{M_{ut}^2} [hN] q_1 [dd]
 +\frac{\kappa_{tu}}{M_{tu}^2} [hN] [dh] \overline{u} \nonumber \\ 
&\sim& 
 \left( \frac{\Lambda_H}{M_{uu}} \right) H q_1 \overline{u} 
 +\left( \frac{\Lambda_H}{M_{ut}} \right)^2 H q_1 \overline{t} 
 +\left( \frac{\Lambda_H}{M_{tu}} \right)^2 H q_3 \overline{u}  
 \label{12generation} \; , 
\end{eqnarray}
where $q_i$ denotes the quark doublet in the i-th generation, 
 $M_{uu}$, $M_{ut}$ and $M_{tu}$ are parameters with dimension one, 
 and we take  $\kappa_{uu} \sim \kappa_{ut} \sim \kappa_{tu} \sim 1$ 
 in the second line.  
It would be able to assume that $M_{uu}\sim M_{ut}\sim M_{tu}$, 
 and the mixing angle between the up and top quarks is given by 
\begin{equation}
 \theta_{ut}\sim \left( \frac{m_u}{m_t} \right)^2 \; . 
\end{equation}  
The same argument is applied to the second generation, 
 and the mixing angle between the charm and top quarks is given by  
 $( {m_c}/{m_t})^2$. 
Note that the flavor mixings are highly suppressed 
 in the up-type quark sector.  

On the other hand, for the down-type quarks and charged leptons, 
 we cannot obtain such relations, 
 since $\overline{H}$ is the elementary particle 
 and no higher dimensional terms are needed to introduce 
 the Yukawa interactions. 
However, we can assume that the mixing angle between 
 the down and bottom quarks is of the order of the (1,3) or (3,1) 
 element of the Kobayashi-Maskawa matrix 
 ($\theta^{KM}_{13} \sim 10^{-3}$ \cite{pdg}), 
 since the mixing angles between the up-type quarks 
 are very small as assumed above. 

Let us discuss the proton life time 
 which is caused by the four-fermion interaction mediated 
 by the colored Higgs and Higgsino. 
The life time is roughly described as 
\begin{equation}
 \tau_p \sim \frac{G_D^{-2}}{m_p^5} \; ,
\end{equation}
 where $G_D$ is the coupling constant of the four-fermion interaction, 
 and $m_p$ is the proton mass.   
The experimental bound is obtained by 
\begin{equation}
G_D < 2 \times 10^{-32} \; \mbox{GeV}^{-2} 
\end{equation}
from $\tau_p > 10^{32}$ yr \cite{pdg}. 

At the tree level, there are decay modes, 
 $p\rightarrow \pi^{0} e^{+} (\mu^{+})$ through lepton mixing 
 and $p \rightarrow \pi^{0} (\tau^{+})^{*}$ ($*$ means off-shell), 
 caused by the four-fermion interaction through the colored Higgs exchange 
 depicted in Fig.6.  
For the mode $p\rightarrow \pi^{0} e^{+} (\mu^{+})$, 
 $G_D$ is roughly estimated as 
\begin{equation}
 G_D \sim \frac{2}{m_D^2} \theta_{ut}^2 \; \theta^{KM}_{13} \; 
 \theta_{e \tau (\mu \tau)} \; ,
\end{equation}
where $\theta_{e \tau (\mu \tau)}$ is the mixing angle 
 between the right-handed electron (muon) and tau, 
 and $m_D$ is the colored Higgs mass (note that 
 $m_D^2= m_{\tilde{q}}^2 + \mu_D^2$ from eqs.(\ref{nseff}) and (18)).    
By substituting the numerical values used in section \ref{sec:ewb} 
 and $m_u=5$ MeV, we obtain 
 $G_D \sim 3 \times 10^{-30}\theta_{e \tau (\mu \tau)}$. 
Therefore, there is no contradiction with the experiments, 
 if $\theta_{e \tau (\mu \tau)} <10^{-2}$. 
In the case of $p \rightarrow \pi^{0} (\tau^{+})^{*}$ mode,      
 $\theta_{e \tau (\mu \tau)}$ is replaced 
 by the factor of the off-shell $\tau$ ``decay'', 
 $G_F m_p^3/m_{\tau} \sim 10^{-5}$. 
We obtain the sufficiently small value, $G_D \sim 3 \times 10^{-35}$. 

At the 1-loop level, the dominant decay modes are 
 $p\rightarrow \pi^{0} e^{+} (\mu^{+})$ and 
 $p\rightarrow \pi^{0} (\tau^{+})^{*}$, 
 which are caused by the four-fermion interaction 
 through the colored Higgsino exchange depicted in Fig.7. 
For the mode  $p \rightarrow \pi^0 e^{+}(\mu^{+})$, 
 we obtain 
\begin{eqnarray}
 G_D &\sim& \frac{\alpha_3}{\pi} 
\frac{1}{m_{\tilde{q}}^2} 
\theta_{ut}^2 \; \theta^{KM}_{13} \; 
 \theta_{e \tau (\mu \tau)} \nonumber \\ 
&\sim & 6\times 10^{-32} \theta_{e \tau (\mu \tau)} \; . 
\end{eqnarray}
Here, note that the dependence on $\mu_D$ and $m_{\lambda_3}$ 
 is absent because of $\mu_D^2 \; , m_{\lambda_3}^2 \ll m_{\tilde{q}}^2$. 
This value of $G_D$ is allowed, 
 if $\theta_{e \tau (\mu \tau)} < 10^{-2}$.  
The mode $p \rightarrow \pi^0 (\tau^{+})^{*}$ is also 
 substantially suppressed, 
 since the angle $\theta_{e \tau (\mu \tau)}$ is replaced 
 by the factor of the off-shell $\tau$ ``decay''.  

>From the following argument, 
 one can expect that the proton decay is 
 sufficiently suppressed in any process. 
Since the baryon number violating interactions exist 
 only in the third generation, 
 the baryon number violating interactions 
 in the first and second generations 
 can be induced only through the flavor mixing 
 between the third and other generations. 
Therefore, in any proton decay process, 
 $G_D$ is always suppressed by the flavor mixing angles. 
The angles in the up-type quark sector can be extremely small 
 as reasonably expected above.  
Moreover, the angles for the flavor mixing 
 of the scalar particles are very small, 
 because of the decoupling effect due to the heavy scalars 
 in the third generation. 
The value of $G_D$ becomes very small by these small mixing angles. 
%
%
\section{The R-axion problem}
\label{sec:axion}

In section \ref{sec:dsb}, we assume that 
 $\langle Z \rangle \neq 0$ and $\langle Y \rangle \neq 0$. 
This means the spontaneous breaking of the global $U(1)_R$ symmetry, 
 and the Nambu-Goldstone boson called the R-axion appears.  
Since the quark superfields have non-zero R-charge, 
 they couple to the R-axion. 
This coupling may cause an astrophysical or a cosmological problem, 
 such as an over cooling of the red giants or 
 a breakdown of the successful predictions from 
 the big bang nucleosynthesis. 
We consider the possibility to avoid the problem. 

Let us introduce $\lambda_X X^3$ term into the tree level superpotential 
 of eq.(\ref{tree}). 
Note that the new term does not disturb the supersymmetry breaking. 
The R-charge of the superfields are changed as follows.  
\begin{center}
\begin{tabular}{ccc} 
 \hspace{1cm} & $ ~~U(1)_R~~ $  \\ 
$ \hat{V} $   &      4/3        \\
$ \hat{Z} $   &      2/3        \\
$ X $         &      2/3        \\
$ Y $         &      2/3        
\end{tabular}
\end{center}
Note that the superfields $Q$ and $\tilde{Q_i}$ must have R-charge, 
 because of the non-zero R-charge of $\hat{V}$. 
Therefore, the new $U(1)_R$ symmetry is explicitly broken by the $SU(2)_S$ 
 supercolor anomaly, and the light R-axion disappears.  
However, the generality of the superpotential is lost by this strategy. 
For example, we cannot introduce $\lambda^{'} Y^3$ term,
 which is allowed by the symmetry, into the superpotential,
 since the term recovers the supersymmetry. 
%
%
\section{summary and comments}
\label{sec:summary}
 
We have constructed a supersymmetric composite model 
 with dynamical supersymmetry breaking.  
The model is based on the gauge group 
 $SU(2)_S \times SU(2)_H \times G_{SM}$. 
There are three building blocks, 
 namely, the dynamical supersymmetry breaking sector, 
 the preon sector recently 
 proposed by Nelson and Strassler, and the standard model sector. 

In the dynamical supersymmetry breaking sector, 
 the non-zero vacuum expectation value of the $F$ component 
 of the gauge-singlet superfield $Z$ arises 
 by the non-perturbative effect of the $SU(2)_S$ supercolor interaction. 
Therefore, supersymmetry is dynamically broken. 
In addition, the scalar component of $Z$ is assumed to have non-zero 
 vacuum expectation value by the effective K\"ahler potential 
 including the radiative corrections due to 
 the massless gauge-singlet superfield $X$. 
The $U(1)_R$ symmetry is spontaneously broken 
 by this vacuum expectation value. 

In the preon sector, the preon superfields are confined 
 into the up-type Higgs 
 in $\bf{5}$ and the superfield in $\bf{10}$ of $SU(5)\supset G_{SM}$ 
 by the non-perturbative effect of the $SU(2)_H$ hypercolor interaction.
The large top Yukawa coupling is dynamically generated, 
 and it is understood as the exchange of the preons 
 in the top quark and Higgs doublet.   

The supersymmetry breaking is mediated to the preon sector 
 and the standard model sector by the radiative corrections. 
The hypercolor gaugino gets the soft breaking mass 
 through the 1-loop diagram in Fig.1, 
 and the scalar preons also get the masses from the 2-loop diagrams 
 in Fig.2.  
The soft breaking masses of the composite scalars, 
 the scalar quarks and lepton and 
 the up-type Higgs boson, originate from the mass of the scalar preons. 
The gauginos in the standard model get the soft breaking masses 
 through the 3-loop diagram in Fig.3. 
It is crucial that the preon superfield $P$ has both charges of 
 the hypercolor and the standard model gauge groups, 
 and it plays a role of the `messenger field' . 
This is a new mechanism to generate the soft breaking masses 
 of the standard model gauginos.  
Below the scale where the hypercolor interaction becomes strong, 
 this mechanism can be understood as the mixing 
 between the gauginos and the massive composite fermions 
 in adjoint representation. 
 
This model predicts the spectrum of the soft breaking masses 
 in the standard model sector. 
All composite scalar fields have the same soft mass, 
 and the masses of the gauginos satisfy the relation 
 of eq.(\ref{relation}). 
All possible soft breaking masses in the Higgs potential are generated 
 by the radiative correction due to the standard model gauge interactions. 
Although the Higgs potential with all the soft breaking masses 
 cannot cause the electroweak symmetry breaking, 
 the breaking is realized by the effect of the large top Yukawa coupling. 
This is the radiative breaking scenario, which originates from 
 the dynamics of the supercolor and the hypercolor interactions. 
We have shown that the electroweak symmetry breaking can occur 
 with experimentally acceptable values of the soft breaking masses 
 in the standard model. 

In our model, there is the colored Higgs which is strongly coupled to 
 quarks and lepton as eq.(\ref{nseff}). 
It seems that the interaction causes too rapid proton decay 
 through the flavor mixing. 
However, in according to eq.(\ref{12generation}), 
 it can be reasonably assumed that 
 the mixing between the third and other generations is highly suppressed. 
As a result, there is no dangerous proton decay. 

There is a potential R-axion problem in the model, 
 since the non-anomalous $U(1)_R$ global symmetry is 
 spontaneously broken by 
 the non-zero vacuum expectation value of the scalar component of $Z$. 
For example, this problem can be avoided 
 by introducing the new term $\lambda_X X^3$  
 into the superpotential of eq.(\ref{tree}), because this term requires 
 to change the charge assignment of $U(1)_R$.   
Since the new $U(1)_R$ symmetry is explicitly broken 
 by the supercolor anomaly,  
 no light R-axion emerges. 
However, the generality of the superpotential is lost. 

Finally, we comment on the flavor changing neutral current between 
 the first and second generations. 
The scalar partners in these generations can get 
 the soft breaking masses by radiative corrections 
 as in Fig.4 and/or Fig.5. 
Since the scalar partners with the same gauge charge are degenerate, 
 the flavor changing neutral current between these two generations 
 is suppressed. 
This feature is the same as the gauge-mediate model \cite{dn}. 
%
\appendix
\section{The Effective K\"ahler Potential}
\label{app:notation}

We discuss the effective K\"ahler potential 
 of the superfield $Z$ generated 
 by the Yukawa coupling $\lambda_{Z}$ in the superpotential 
 of eq.(\ref{tree}). 
We calculate the infinite series of the 1-loop graphs, 
 and obtain the effective K\"ahler potential as 
\begin{equation}
K_{eff}= Z^{\dagger}Z 
 \left[1 - \frac{\lambda^2_Z}{8 \pi^2} \ln 
 \left( \frac{\lambda_Z^2 Z^{\dagger} Z} {\mu^2}
 \right)
 \right]  \label{kahler}\; ,
\end{equation}
where $\mu$ is the scale introduced 
 by the wave function renormalization of the superfield $Z$. 
Then, the vacuum energy is given by 
\begin{equation}
E_{vac}=  \left| \frac{\partial W}{\partial Z} \right|^2  \Bigg/
 \left[ 1-\frac{\lambda^2_Z}{8 \pi^2} \left\{ 
 2+ \ln \left( \frac{\lambda_Z^2 Z^{\dagger} Z }{\mu^2}
 \right) \right\} \right]_{Z= \langle Z \rangle } \; .  
\end{equation}
One may expect that the vacuum is realized at $\langle Z \rangle= 0 $, 
 since the denominator diverges and $E_{vac}$ vanishes.  
However, the 1-loop approximation cannot be valid near the origin, 
 since the logarithm becomes too large in eq.(\ref{kahler}).  
Although the complete calculation seems to be impossible like 
 in the non-supersymmetric 
 massless $\lambda \phi^4$ theory discussed by Coleman and Weinberg, 
 we expect that $\langle Z \rangle \neq 0$ is realized. 
The appearance of the non-zero vacuum expectation value  
 seems to be a common mechanism to solve the problem 
 of the infrared divergence 
 connected with the singularity of the effective potential at the origin. 
Indeed, Coleman and Weinberg have shown that such mechanism really works  
 in the  non-supersymmetric massless $\lambda \phi^4$ theory with $U(1)$ 
 gauge interaction in their original paper.  


\begin{figure}
\caption{The 1-loop diagram for the soft breaking mass 
 of the hypercolor gaugino. 
 The solid and dotted internal lines denote the propagators 
 of $\psi_{V_i}$ and $V_i$ ($i=1,2$), respectively. 
 The crosses denote the insertions of $F$ and $m$ 
 for the internal boson and fermion lines, 
 respectively. }
\end{figure}
\begin{figure}
\caption{The 2-loop diagrams for the soft breaking masses 
 of the scalar preons. 
 The solid and dotted internal lines denote the propagators 
 of $\psi_{V_i}$ and $V_i$ ($i=1,2$), respectively.
 The dashed internal line denotes the propagator of the scalar preon 
 $\tilde{P}$ or $\tilde{N}$. 
 The wavy line denotes the hypercolor gauge boson propagator.  
 The propagator of the hypercolor gaugino is denoted by the wavy line 
 with the solid line. 
 The crosses denote the insertions of $F$.} 
\end{figure}
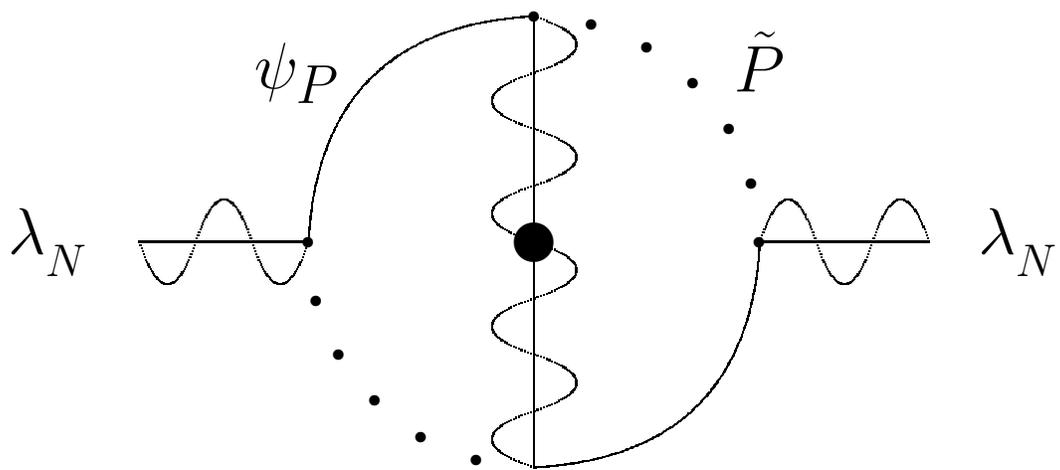
\begin{figure}
\caption{The 3-loop diagram for the soft breaking masses 
 of the standard model gauginos. 
 The solid and dotted internal lines denote the propagators 
 of the preon $\psi_P$ and the scalar 
 preon $\tilde{P}$, respectively.
 The internal line of the hypercolor gaugino is denoted by the wavy line 
 with the solid line.  
 The dot on the propagator of the hypercolor gaugino denotes 
 the 1-loop diagram in Fig.1.}
\end{figure}
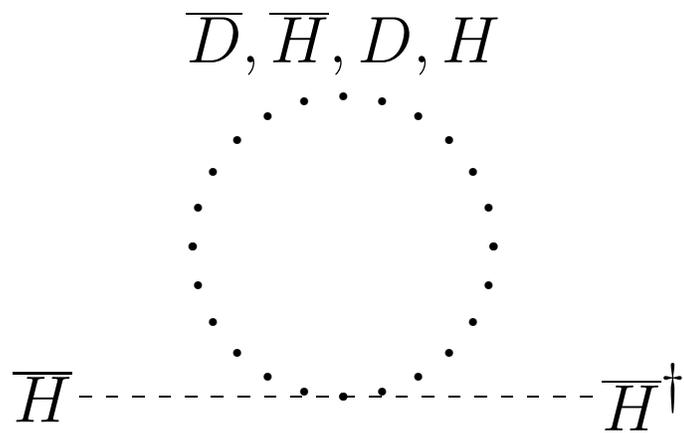
\begin{figure}
\caption{The 1-loop diagram for the soft breaking mass 
 of the down-type Higgs boson.  
 The dotted internal line denotes the propagators of $\overline{D}$, 
 $\overline{H}$, $D$, and $H$. 
 It is not needed to consider the corrections of the scalar quarks 
 and leptons, 
 because their contributions cancel out each other.}  
\end{figure}
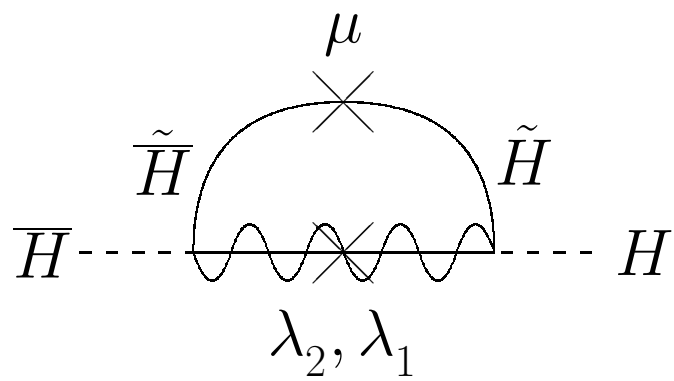
\begin{figure}
\caption{The 1-loop diagram to generate the soft breaking term 
 $B \overline{H} H + h.c.$ in the Higgs potential.  
 The solid internal lines denote the propagators 
 of the Higgsinos $\tilde{\overline{H}}$ or $\tilde{H}$. 
 The wavy line with the solid line denotes the propagator 
 of the $SU(2)_L$ gaugino or the $U(1)_Y$ gaugino.  
 Since the contribution of the $U(1)_Y$ gaugino loop 
 is far smaller than that of 
 the $SU(2)_L$ gaugino, it can be neglected in eq.(21).}
\end{figure}
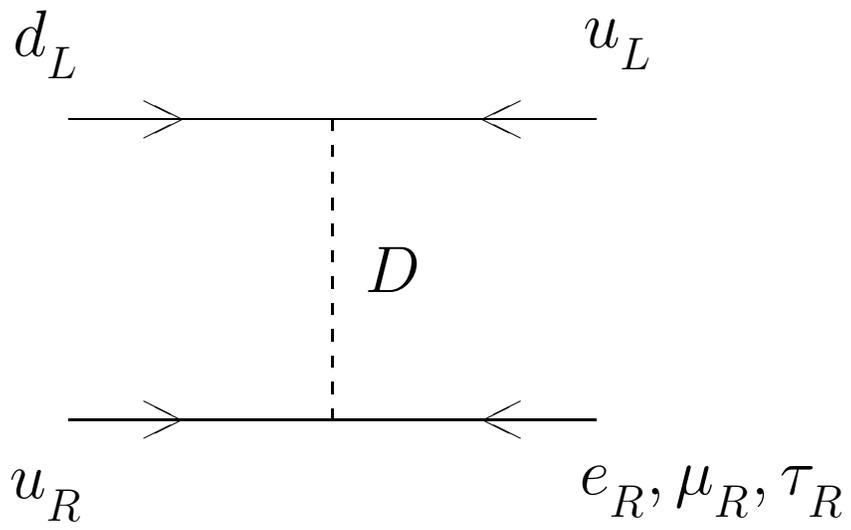
\begin{figure}
\caption{The four-fermion interaction caused by 
 the colored Higgs exchange.  
 $D$ denotes the colored Higgs. 
 } 
\end{figure}
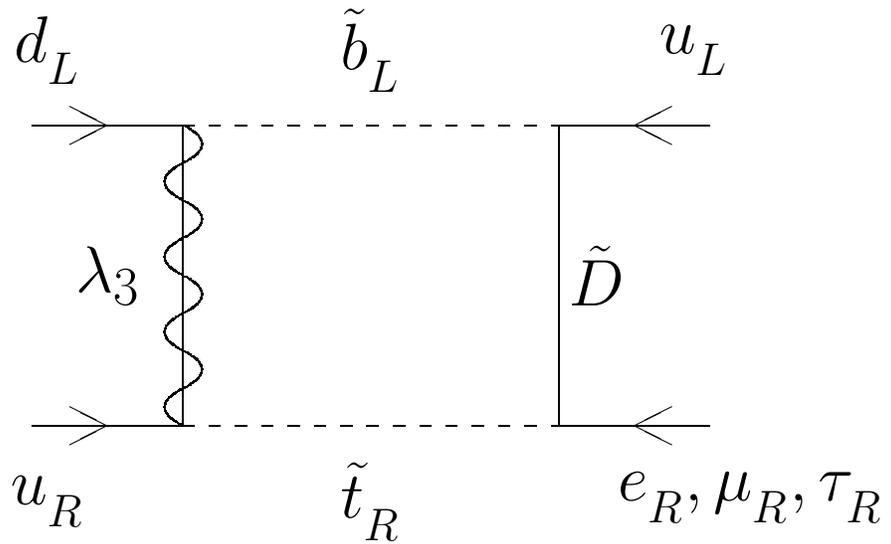
\begin{figure}
\caption{The four-fermion interactions caused 
 by the colored Higgsino exchange. 
 $\tilde{D}$ denotes the colored Higgsino.  
 The diagram by replacement 
 $d_{{}_L} \leftrightarrow u_{{}_L}$ 
 and $\tilde{b}_{{}_L} \rightarrow \tilde{t}_{{}_L}$ 
 is also possible.                               
 }
\end{figure}
\newpage
%
\begin{picture}(160,200)
\put(70,200){\begin{minipage}{10cm}
\unitlength=0.15mm
\begin{picture}(1000,1000)(-350,200)
\bezier{80}(-350,0)(-325,-75)(-300,0)
\bezier{80}(-300,0)(-275, 75)(-250,0)
\bezier{80}(-250,0)(-225,-75)(-200,0)
\bezier{80}( 300,0)( 275,-75)( 250,0)
\bezier{80}( 250,0)( 225, 75)( 200,0)
\bezier{80}( 350,0)( 325, 75)( 300,0)
\put(-350, 0){\line(1,0){700}}
\put(0,200){\line(1,2){20}}
\put(0,200){\line(1,-2){20}}
\put(0,200){\line(-1,2){20}}
\put(0,200){\line(-1,-2){20}}
\put(-50,-120){\makebox(100,100){\huge{$m$}}}
\put(-50,240){\makebox(100,100){\huge{$F$}}}
\put(-480,-50){\makebox(100,100){\huge{$\lambda_H$}}}
\put(380,-50){\makebox(100,100){\huge{$\lambda_H$}}}
\put(150,150){\makebox(100,100){\huge{$V_2$}}}
\put(-250,150){\makebox(100,100){\huge{$V_1$}}}
\put(120,-150){\makebox(100,100){\huge{$\psi_{{}_{V_2}}$}}}
\put(-180,-150){\makebox(100,100){\huge{$\psi_{{}_{V_1}}$}}}
\put(0,0){\line(1,2){20}}
\put(0,0){\line(1,-2){20}}
\put(0,0){\line(-1,2){20}}
\put(0,0){\line(-1,-2){20}}
\put(200,0){\circle*{10}}
\put(193,52){\circle*{10}}
\put(173,100){\circle*{10}}
\put(141,141){\circle*{10}}
\put(100,173){\circle*{10}}
\put(52,193){\circle*{10}}
\put(0,200){\circle*{10}}
\put(-200,0){\circle*{10}}
\put(-193,52){\circle*{10}}
\put(-173,100){\circle*{10}}
\put(-141,141){\circle*{10}}
\put(-100,173){\circle*{10}}
\put(-52,193){\circle*{10}}
\put(-200,-320){\makebox(400,100){\Huge{Fig.1}}}
\end{picture}
\end{minipage}}
\end{picture}
\newpage
%
%
\begin{picture}(170,100)
\put(50,50){\begin{minipage}[t]{10cm}
\unitlength=0.07mm
\begin{picture}(1000,1000)(50,550)
\bezier{50}(250,0)(346,47)(240,62)
\bezier{50}(240,62)(161,57)(212,118)
\bezier{50}(212,118)(262,212)(168,162)
\bezier{50}(168,162)(107,111)(112,190)
\bezier{50}(112,190)(110,260)(75,220)
\bezier{50}(-250,0)(-346,47)(-240,62)
\bezier{50}(-240,62)(-161,57)(-212,118)
\bezier{50}(-212,118)(-262,212)(-168,162)
\bezier{50}(-168,162)(-107,111)(-112,190)
\bezier{50}(-112,190)(-110,260)(-75,220)
\multiput(-350,0)(35,0){20}{\line(1,0){15}}
\put(0,220){\oval(150,150)}
\put(0,220){\line(1,1){55}}
\put(0,220){\line(-1,-1){55}}
\put(0,220){\line(1,-1){55}}
\put(0,220){\line(-1,1){55}}
\put(-100,-270){\makebox(200,150){\Huge{(a)}}}
\end{picture}
\end{minipage}}
\put(150,50){\begin{minipage}[t]{10cm}
\unitlength=0.07mm
\begin{picture}(1000,1000)(-300,350)
\bezier{50}(145,39)(217,125)(106,106)
\bezier{50}(106,106)(70,110)(75,150)
\bezier{50}(-106,106)(-70,110)(-75,150)
\bezier{50}(145,39)(100,0)(145,-39)
\bezier{50}(-145,39)(-217,125)(-106,106)
\bezier{50}(145,-39)(217,-125)(106,-106)
\bezier{50}(106,-106)(50,-87)(39,-145)
\bezier{50}(39,-145)(0,-250)(-39,-145)
\bezier{50}(145,-39)(100,0)(145,39)
\bezier{50}(-145,-39)(-217,-125)(-106,-106)
\bezier{50}(-106,-106)(-50,-87)(-39,-145)
\bezier{50}(-39,-145)(-0,-250)(39,-145)
\bezier{50}(-145,-39)(-100,-0)(-145,39)
\multiput(-350,-200)(35,0){20}{\line(1,0){15}}
\put(0,150){\oval(150,150)}
\put(0,150){\line(1,1){55}}
\put(0,150){\line(-1,-1){55}}
\put(0,150){\line(1,-1){55}}
\put(0,150){\line(-1,1){55}}
\put(-100,-470){\makebox(200,150){\Huge{(b)}}}
\put(-100,420){\makebox(200,200){\Huge{Fig.2}}}
\end{picture}
\end{minipage}}
\put(250,50){\begin{minipage}[t]{10cm}
\unitlength=0.07mm
\begin{picture}(1000,1000)(-650,550)
\bezier{50}(200,0)(125,25)(200,50)
\bezier{50}(200,50)(275,75)(200,100)
\bezier{50}(200,100)(125,125)(200,150)
\bezier{50}(200,150)(275,175)(200,200)
\put(200,0){\line(0,1){200}}
\bezier{50}(-200,0)(-275,25)(-200,50)
\bezier{50}(-200,50)(-125,75)(-200,100)
\bezier{50}(-200,100)(-275,125)(-200,150)
\bezier{50}(-200,150)(-125,175)(-200,200)
\put(-200,0){\line(0,1){200}}
\put(-200,200){\line(1,0){400}}
\multiput(-360,0)(35,0){21}{\line(1,0){15}}
\put(200,200){\circle*{10}}
\put(193,252){\circle*{10}}
\put(173,300){\circle*{10}}
\put(141,341){\circle*{10}}
\put(100,373){\circle*{10}}
\put(51,393){\circle*{10}}
\put(0,400){\circle*{10}}
\put(-200,200){\circle*{10}}
\put(-193,252){\circle*{10}}
\put(-173,300){\circle*{10}}
\put(-141,341){\circle*{10}}
\put(-100,373){\circle*{10}}
\put(-51,393){\circle*{10}}
\put(100,373){\line(1,2){30}}
\put(100,373){\line(-1,2){30}}
\put(100,373){\line(1,-2){30}}
\put(100,373){\line(-1,-2){30}}
\put(-100,373){\line(1,2){30}}
\put(-100,373){\line(-1,2){30}}
\put(-100,373){\line(1,-2){30}}
\put(-100,373){\line(-1,-2){30}}
\put(-100,-270){\makebox(200,150){\Huge{(c)}}}
\end{picture}
\end{minipage}}
\put(50,100){\begin{minipage}{10cm}
\unitlength=0.07mm
\begin{picture}(1000,1000)(50,1100)
\multiput(-350,0)(35,0){20}{\line(1,0){15}}
\put(200,0){\circle*{10}}
\put(193,52){\circle*{10}}
\put(173,100){\circle*{10}}
\put(141,141){\circle*{10}}
\put(100,173){\circle*{10}}
\put(51,193){\circle*{10}}
\put(0,200){\circle*{10}}
\put(-200,0){\circle*{10}}
\put(-193,52){\circle*{10}}
\put(-173,100){\circle*{10}}
\put(-141,141){\circle*{10}}
\put(-100,173){\circle*{10}}
\put(-51,193){\circle*{10}}
\put(193,-52){\circle*{10}}
\put(173,-100){\circle*{10}}
\put(141,-141){\circle*{10}}
\put(100,-173){\circle*{10}}
\put(51,-193){\circle*{10}}
\put(0,-200){\circle*{10}}
\put(-200,0){\circle*{10}}
\put(-193,-52){\circle*{10}}
\put(-173,-100){\circle*{10}}
\put(-141,-141){\circle*{10}}
\put(-100,-173){\circle*{10}}
\put(-51,-193){\circle*{10}}
\put(0,200){\line(1,2){40}}
\put(0,200){\line(-1,2){40}}
\put(0,200){\line(1,-2){40}}
\put(0,200){\line(-1,-2){40}}
\put(0,-200){\line(1,2){40}}
\put(0,-200){\line(-1,2){40}}
\put(0,-200){\line(1,-2){40}}
\put(0,-200){\line(-1,-2){40}}
\put(-200,-550){\makebox(400,150){\Huge{(d-1)}}}
\end{picture}
\end{minipage}}
\put(150,100){\begin{minipage}{15cm}
\unitlength=0.07mm
\begin{picture}(1000,1000)(-300,1100)
\multiput(-350,0)(35,0){20}{\line(1,0){15}}
\put(200,0){\circle*{10}}
\put(193,52){\circle*{10}}
\put(173,100){\circle*{10}}
\put(141,141){\circle*{10}}
\put(100,173){\circle*{10}}
\put(51,193){\circle*{10}}
\put(0,200){\circle*{10}}
\put(-200,0){\circle*{10}}
\put(-193,52){\circle*{10}}
\put(-173,100){\circle*{10}}
\put(-141,141){\circle*{10}}
\put(-100,173){\circle*{10}}
\put(-51,193){\circle*{10}}
\put(193,-52){\circle*{10}}
\put(173,-100){\circle*{10}}
\put(141,-141){\circle*{10}}
\put(100,-173){\circle*{10}}
\put(51,-193){\circle*{10}}
\put(0,-200){\circle*{10}}
\put(-200,0){\circle*{10}}
\put(-193,-52){\circle*{10}}
\put(-173,-100){\circle*{10}}
\put(-141,-141){\circle*{10}}
\put(-100,-173){\circle*{10}}
\put(-51,-193){\circle*{10}}
\put(100,173){\line(1,2){40}}
\put(100,173){\line(-1,2){40}}
\put(100,173){\line(1,-2){40}}
\put(100,173){\line(-1,-2){40}}
\put(-100,173){\line(1,2){40}}
\put(-100,173){\line(-1,2){40}}
\put(-100,173){\line(1,-2){40}}
\put(-100,173){\line(-1,-2){40}}
\put(-100,-550){\makebox(200,150){\Huge{(d-2)}}}
\end{picture}
\end{minipage}}
\put(250,100){\begin{minipage}{15cm}
\unitlength=0.07mm
\begin{picture}(1000,1000)(-650,1100)
\multiput(-350,0)(35,0){20}{\line(1,0){15}}
\put(200,0){\circle*{10}}
\put(193,52){\circle*{10}}
\put(173,100){\circle*{10}}
\put(141,141){\circle*{10}}
\put(100,173){\circle*{10}}
\put(51,193){\circle*{10}}
\put(0,200){\circle*{10}}
\put(-200,0){\circle*{10}}
\put(-193,52){\circle*{10}}
\put(-173,100){\circle*{10}}
\put(-141,141){\circle*{10}}
\put(-100,173){\circle*{10}}
\put(-51,193){\circle*{10}}
\put(193,-52){\circle*{10}}
\put(173,-100){\circle*{10}}
\put(141,-141){\circle*{10}}
\put(100,-173){\circle*{10}}
\put(51,-193){\circle*{10}}
\put(0,-200){\circle*{10}}
\put(-200,0){\circle*{10}}
\put(-193,-52){\circle*{10}}
\put(-173,-100){\circle*{10}}
\put(-141,-141){\circle*{10}}
\put(-100,-173){\circle*{10}}
\put(-51,-193){\circle*{10}}
\put(100,-173){\line(1,2){40}}
\put(100,-173){\line(-1,2){40}}
\put(100,-173){\line(1,-2){40}}
\put(100,-173){\line(-1,-2){40}}
\put(-100,-173){\line(1,2){40}}
\put(-100,-173){\line(-1,2){40}}
\put(-100,-173){\line(1,-2){40}}
\put(-100,-173){\line(-1,-2){40}}
\put(-100,-550){\makebox(200,150){\Huge{(d-3)}}}
\end{picture}
\end{minipage}}
%
\put(60,-200){\begin{minipage}{10cm}
\unitlength=0.06mm
\begin{picture}(1000,1000)(100,800)
\bezier{50}(-250,0)(-225,-75)(-200,0)
\bezier{50}(-200,0)(-175, 75)(-150,0)
\bezier{50}(-150,0)(-125,-75)(-90,0)
\bezier{50}( 250,0)( 225, 75)( 200,0)
\bezier{50}( 200,0)( 175,-75)( 150,0)
\bezier{50}( 150,0)( 125, 75)( 90,0)
\put(0,0){\oval(180,180)}
\put(350,-50){\makebox(200,100){\Huge{$\equiv$}}}
\put(0,0){\line(1,1){60}}
\put(0,0){\line(1,-1){60}}
\put(0,0){\line(-1,1){60}}
\put(0,0){\line(-1,-1){60}}
\end{picture}
\end{minipage}}
\put(130,-200){\begin{minipage}{15cm}
\unitlength=0.05mm
\begin{picture}(1000,1000)(-500,1000)
\put(200,0){\circle*{10}}
\put(193,52){\circle*{10}}
\put(173,100){\circle*{10}}
\put(141,141){\circle*{10}}
\put(100,173){\circle*{10}}
\put(51,193){\circle*{10}}
\put(0,200){\circle*{10}}
\put(-200,0){\circle*{10}}
\put(-193,52){\circle*{10}}
\put(-173,100){\circle*{10}}
\put(-141,141){\circle*{10}}
\put(-100,173){\circle*{10}}
\put(-51,193){\circle*{10}}
\put(193,-52){\circle*{10}}
\put(173,-100){\circle*{10}}
\put(141,-141){\circle*{10}}
\put(100,-173){\circle*{10}}
\put(51,-193){\circle*{10}}
\put(0,-200){\circle*{10}}
\put(-200,0){\circle*{10}}
\put(-193,-52){\circle*{10}}
\put(-173,-100){\circle*{10}}
\put(-141,-141){\circle*{10}}
\put(-100,-173){\circle*{10}}
\put(-51,-193){\circle*{10}}
\bezier{50}(275,-225)(250,-300)(225,-225)
\bezier{50}(225,-225)(200,-150)(175,-225)
\bezier{50}(175,-225)(150,-300)(125,-225)
\bezier{50}(125,-225)(100,-150)(75,-225)
\bezier{50}(75,-225)(50,-300)(25,-225)
\bezier{50}(-275,-225)(-250,-300)(-225,-225)
\bezier{50}(-225,-225)(-200,-150)(-175,-225)
\bezier{50}(-175,-225)(-150,-300)(-125,-225)
\bezier{50}(-125,-225)(-100,-150)(-75,-225)
\bezier{50}(-75,-225)(-50,-300)(-25,-225)
\bezier{50}(-25,-225)( 0,-170)(25,-225)
\bezier{50}(25,-225)( 50,-300)(75,-225)
\put(100,173){\line(1,2){40}}
\put(100,173){\line(-1,2){40}}
\put(100,173){\line(1,-2){40}}
\put(100,173){\line(-1,-2){40}}
\put(-100,173){\line(1,2){40}}
\put(-100,173){\line(-1,2){40}}
\put(-100,173){\line(1,-2){40}}
\put(-100,173){\line(-1,-2){40}}
\end{picture}
\end{minipage}}
\end{picture}
\put(220,-200){\begin{minipage}{10cm}
\unitlength=0.05mm
\begin{picture}(1000,1000)(200,1000)
\put(200,0){\circle*{10}}
\put(193,52){\circle*{10}}
\put(173,100){\circle*{10}}
\put(141,141){\circle*{10}}
\put(100,173){\circle*{10}}
\put(51,193){\circle*{10}}
\put(0,200){\circle*{10}}
\put(-200,0){\circle*{10}}
\put(-193,52){\circle*{10}}
\put(-173,100){\circle*{10}}
\put(-141,141){\circle*{10}}
\put(-100,173){\circle*{10}}
\put(-51,193){\circle*{10}}
\put(193,-52){\circle*{10}}
\put(173,-100){\circle*{10}}
\put(141,-141){\circle*{10}}
\put(100,-173){\circle*{10}}
\put(51,-193){\circle*{10}}
\put(0,-200){\circle*{10}}
\put(-200,0){\circle*{10}}
\put(-193,-52){\circle*{10}}
\put(-173,-100){\circle*{10}}
\put(-141,-141){\circle*{10}}
\put(-100,-173){\circle*{10}}
\put(-51,-193){\circle*{10}}
\bezier{50}(350,0)(325,75)(300,0)
\bezier{50}(300,0)(275,-75)(250,0)
\bezier{50}(250,0)(225,75)(200,0)
\bezier{50}(-350,0)(-325,-75)(-300,0)
\bezier{50}(-300,0)(-275,75)(-250,0)
\bezier{50}(-250,0)(-225,-75)(-200,0)
%
\put(-600,-50){\makebox(100,100){\huge{$+$}}} 
\put(0,200){\line(1,2){40}}
\put(0,200){\line(1,-2){40}}
\put(0,200){\line(-1,2){40}}
\put(0,200){\line(-1,-2){40}}
\put(0,-200){\line(1,2){40}}
\put(0,-200){\line(1,-2){40}}
\put(0,-200){\line(-1,2){40}}
\put(0,-200){\line(-1,-2){40}}
\end{picture}
\end{minipage}}
\put(0,-300){\begin{minipage}{15cm}
\unitlength=0.05mm
\begin{picture}(1000,1000)(-200,1000) 
\put(200,0){\circle*{10}}
\put(193,52){\circle*{10}}
\put(173,100){\circle*{10}}
\put(141,141){\circle*{10}}
\put(100,173){\circle*{10}}
\put(51,193){\circle*{10}}
\put(0,200){\circle*{10}}
\put(-200,0){\circle*{10}}
\put(-193,52){\circle*{10}}
\put(-173,100){\circle*{10}}
\put(-141,141){\circle*{10}}
\put(-100,173){\circle*{10}}
\put(-51,193){\circle*{10}}
\put(193,-52){\circle*{10}}
\put(173,-100){\circle*{10}}
\put(141,-141){\circle*{10}}
\put(100,-173){\circle*{10}}
\put(51,-193){\circle*{10}}
\put(0,-200){\circle*{10}}
\put(-200,0){\circle*{10}}
\put(-193,-52){\circle*{10}}
\put(-173,-100){\circle*{10}}
\put(-141,-141){\circle*{10}}
\put(-100,-173){\circle*{10}}
\put(-51,-193){\circle*{10}}
\bezier{50}(350,0)(325,75)(300,0)
\bezier{50}(300,0)(275,-75)(250,0)
\bezier{50}(250,0)(225,75)(200,0)
\bezier{50}(-350,0)(-325,-75)(-300,0)
\bezier{50}(-300,0)(-275,75)(-250,0)
\bezier{50}(-250,0)(-225,-75)(-200,0)
\put(100,173){\line(1,2){40}}
\put(100,173){\line(-1,2){40}}
\put(100,173){\line(1,-2){40}}
\put(100,173){\line(-1,-2){40}}
\put(-100,173){\line(1,2){40}}
\put(-100,173){\line(-1,2){40}}
\put(-100,173){\line(1,-2){40}}
\put(-100,173){\line(-1,-2){40}}
\put(-600,-50){\makebox(100,100){\Huge{$+$}}}
\end{picture}
\end{minipage}}
\put(220,-300){\begin{minipage}{15cm}
\unitlength=0.05mm
\begin{picture}(1000,1000)(200,1000)
\put(200,0){\circle*{10}}
\put(193,52){\circle*{10}}
\put(173,100){\circle*{10}}
\put(141,141){\circle*{10}}
\put(100,173){\circle*{10}}
\put(51,193){\circle*{10}}
\put(0,200){\circle*{10}}
\put(-200,0){\circle*{10}}
\put(-193,52){\circle*{10}}
\put(-173,100){\circle*{10}}
\put(-141,141){\circle*{10}}
\put(-100,173){\circle*{10}}
\put(-51,193){\circle*{10}}
\put(193,-52){\circle*{10}}
\put(173,-100){\circle*{10}}
\put(141,-141){\circle*{10}}
\put(100,-173){\circle*{10}}
\put(51,-193){\circle*{10}}
\put(0,-200){\circle*{10}}
\put(-200,0){\circle*{10}}
\put(-193,-52){\circle*{10}}
\put(-173,-100){\circle*{10}}
\put(-141,-141){\circle*{10}}
\put(-100,-173){\circle*{10}}
\put(-51,-193){\circle*{10}}
\bezier{50}(350,0)(325,75)(300,0)
\bezier{50}(300,0)(275,-75)(250,0)
\bezier{50}(250,0)(225,75)(200,0)
\bezier{50}(-350,0)(-325,-75)(-300,0)
\bezier{50}(-300,0)(-275,75)(-250,0)
\bezier{50}(-250,0)(-225,-75)(-200,0)
\put(100,-173){\line(1,2){40}}
\put(100,-173){\line(-1,2){40}}
\put(100,-173){\line(1,-2){40}}
\put(100,-173){\line(-1,-2){40}}
\put(-100,-173){\line(1,2){40}}
\put(-100,-173){\line(-1,2){40}}
\put(-100,-173){\line(1,-2){40}}
\put(-100,-173){\line(-1,-2){40}}
\put(-600,-50){\makebox(100,100){\Huge{$+$}}}
\end{picture}
\end{minipage}}
\newpage
%
%
\begin{picture}(160,200)
\put(70,0){\begin{minipage}[t]{10cm}
\unitlength=0.15mm
\begin{picture}(1000,1000)(-350,100)
\bezier{50}(-350,0)(-325,-75)(-300,0)
\bezier{50}(-300,0)(-275, 75)(-250,0)
\bezier{50}(-250,0)(-225,-75)(-200,0)
\bezier{50}( 300,0)( 275,-75)( 250,0)
\bezier{50}( 250,0)( 225, 75)( 200,0)
\bezier{50}( 350,0)( 325, 75)( 300,0)
\put(-350,0){\line(1,0){150}}
\put(200,0){\line(1,0){150}}
\bezier{50}(0,0)(-75,25)(0,50)
\bezier{50}(0,50)(75,75)(0,100)
\bezier{50}(0,100)(-75,125)(0,150)
\bezier{50}(0,150)(75,175)(0,200)
\bezier{50}(0,0)(75,-25)(0,-50)
\bezier{50}(0,-50)(-75,-75)(0,-100)
\bezier{50}(0,-100)(75,-125)(0,-150)
\bezier{50}(0,-150)(-75,-175)(0,-200)
\put(0,-200){\line(0,1){400}}
\bezier{200}(200,0)(190,-190)(0,-200)
\bezier{200}(-200,0)(-190,190)(0,200)
\put(200,0){\circle*{10}}
\put(193,52){\circle*{10}}
\put(173,100){\circle*{10}}
\put(141,141){\circle*{10}}
\put(100,173){\circle*{10}}
\put(51,193){\circle*{10}}
\put(0,200){\circle*{10}}
\put(-200,-0){\circle*{10}}
\put(-193,-52){\circle*{10}}
\put(-173,-100){\circle*{10}}
\put(-141,-141){\circle*{10}}
\put(-100,-173){\circle*{10}}
\put(-51,-193){\circle*{10}}
\put(-480,-50){\makebox(100,100){\huge{$\lambda_{{}_N}$}}}
\put(380,-50){\makebox(100,100){\huge{$\lambda_{{}_N}$}}}
\put(150,110){\makebox(100,100){\huge{$\tilde{P}$}}}
\put(-260,100){\makebox(100,100){\huge{$\psi_P$}}}
\put(0,0){\circle*{60}}
\put(-200,-480){\makebox(400,100){\Huge{Fig.3}}}
\end{picture}
\end{minipage}}
\end{picture}
\newpage
%
%
\begin{picture}(160,200)
\put(80,180){\begin{minipage}[t]{10cm}
\unitlength=0.1mm
\begin{picture}(1000,1000)(-500,700)
\multiput(-350,0)(35,0){20}{\line(1,0){15}}
\put(-450,-50){\makebox(100,100){\huge{$\overline{H}$}}}
\put( 350,-50){\makebox(100,100){\huge{${\overline{H}}^{\dagger}$}}}
\put( -200,420){\makebox(400,100){\huge{${\overline{D}},
{\overline{H}},D,H$}}}
\put(200,200){\circle*{10}}
\put(193,252){\circle*{10}}
\put(173,300){\circle*{10}}
\put(141,341){\circle*{10}}
\put(100,373){\circle*{10}}
\put(52,393){\circle*{10}}
\put(0,400){\circle*{10}}
\put(-200,200){\circle*{10}}
\put(-193,252){\circle*{10}}
\put(-173,300){\circle*{10}}
\put(-141,341){\circle*{10}}
\put(-100,373){\circle*{10}}
\put(-52,393){\circle*{10}}
\put(200,200){\circle*{10}}
\put(193,148){\circle*{10}}
\put(173,100){\circle*{10}}
\put(141,59){\circle*{10}}
\put(100,27){\circle*{10}}
\put(52,7){\circle*{10}}
\put(0,0){\circle*{10}}
\put(-200,200){\circle*{10}}
\put(-193,148){\circle*{10}}
\put(-173,100){\circle*{10}}
\put(-141,59){\circle*{10}}
\put(-100,27){\circle*{10}}
\put(-52,7){\circle*{10}}
\put(-0,0){\circle*{10}}
\put( -200,-250){\makebox(400,100){\Huge{Fig.4}}}
\end{picture}
\end{minipage}}
%
%
\put(80,-100){\begin{minipage}[b]{10cm}
\unitlength=0.1mm
\begin{picture}(1000,1000)(-500,700)
\bezier{50}(-200,0)(-175,-75)(-150,0)
\bezier{50}(-150,0)(-125, 75)(-100,0)
\bezier{50}(-100,0)(-75,-75)(-50,0)
\bezier{50}(-50,0)(-25,75)(0,0)
\bezier{50}(200,0)(175,75)(150,0)
\bezier{50}(150,0)(125,-75)(100,0)
\bezier{50}(100,0)(75,75)(50,0)
\bezier{50}(50,0)(25,-75)(0,0)
\bezier{200}(-200,0)(-200,200)(0, 200)
\bezier{200}( 200,0)( 200,200)(0, 200)
\put(-200, 0){\line(1,0){400}}
\multiput(-350,0)(35,0){20}{\line(1,0){15}}
\put(-50,240){\makebox(100,100){\huge{$\mu $}}}
\put(-100,-170){\makebox(200,100){\huge{$\lambda_{{}_2},\lambda_{{}_1} $}}}
\put(-450,-50){\makebox(100,100){\huge{$\overline{H}$}}}
\put( 350,-50){\makebox(100,100){\huge{$H$}}}
\put(-290,70){\makebox(100,100){\huge{$\tilde{\overline{H}}$}}}
\put(190,80){\makebox(100,100){\huge{$\tilde{H}$}}}
\put(0,200){\line(1,1){40}}
\put(0,200){\line(1,-1){40}}
\put(0,200){\line(-1,1){40}}
\put(0,200){\line(-1,-1){40}}
\put(0,0){\line(1,1){40}}
\put(0,0){\line(1,-1){40}}
\put(0,0){\line(-1,1){40}}
\put(0,0){\line(-1,-1){40}}
\put( -200,-380){\makebox(400,100){\Huge{Fig.5}}}
\end{picture}
\end{minipage}}
\end{picture}
\newpage
%
%
%
\begin{picture}(160,50)
\put(80,20){\begin{minipage}[t]{10cm}
\unitlength=0.1mm
\begin{picture}(1000,1000)(-500,500)
\multiput(0,0)(0,35){12}{\line(0,1){15}}
\put(-430,-150){\makebox(100,100){\huge{$u_{{}_R}$}}} 
\put(-430, 450){\makebox(100,100){\huge{$d_{{}_L}$}}}
\put(330, 450){\makebox(100,100){\huge{$u_{{}_L}$}}}
\put(430, -170){\makebox(150,150){\huge{$e_{{}_R}, \mu_{{}_R}, 
\tau_{{}_R}$}}}
\put(30, 150){\makebox(100,100){\huge{$D$}}}
\put(-350,400){\line(1,0){700}}
\put(-350,0){\line(1,0){700}}
\put(-200,0){\line(-2,1){50}}
\put(-200,0){\line(-2,-1){50}}
\put(200,0){\line(2,1){50}}
\put(200,0){\line(2,-1){50}}
\put(-200,400){\line(-2,1){50}}
\put(-200,400){\line(-2,-1){50}}
\put(200,400){\line(2,1){50}}
\put(200,400){\line(2,-1){50}}
\put(-200,-400){\makebox(400,100){\Huge{Fig.6}}}
\end{picture}
\end{minipage}}
%
%
\put(370,-340){\begin{minipage}[b]{10cm}
\unitlength=0.1mm
\begin{picture}(1000,1000)(500,500)
\multiput(-250,0)(35,0){15}{\line(1,0){15}}
\multiput(-250,400)(35,0){15}{\line(1,0){15}}
\put(-480,-150){\makebox(100,100){\huge{$u_{{}_R}$}}} 
\put(-480, 450){\makebox(100,100){\huge{$d_{{}_L}$}}}
\put(380, 450){\makebox(100,100){\huge{$u_{{}_L}$}}}
\put(430, -170){\makebox(150,150){\huge{$e_{{}_R}, 
\mu_{{}_R}, \tau_{{}_R}$}}}
\put(-50,450){\makebox(100,100){\huge{$\tilde{b}_{{}_L}$}}}
\put(-50, -150){\makebox(100,100){\huge{$\tilde{t}_{{}_R}$}}}
\put(-400, 150){\makebox(100,100){\huge{$\lambda_3$}}}
\put(250, 150){\makebox(100,100){\huge{$\tilde{D}$}}}
\put(-250,400){\line(-1,0){200}}
\put(-250,-0){\line(-1,0){200}}
\put(250,400){\line(1,0){200}}
\put(250,0){\line(1,0){200}}
\put(-250,0){\line(0,1){400}}
\put(250,0){\line(0,1){400}}
\put(-350,0){\line(-2,1){50}}
\put(-350,0){\line(-2,-1){50}}
\put(350,0){\line(2,1){50}}
\put(350,0){\line(2,-1){50}}
\put(-350,400){\line(-2,1){50}}
\put(-350,400){\line(-2,-1){50}}
\put(350,400){\line(2,1){50}}
\put(350,400){\line(2,-1){50}}
\bezier{50}(-250,0)(-300,25)(-250,50)
\bezier{50}(-250,50)(-200,75)(-250,100)
\bezier{50}(-250,100)(-300,125)(-250,150)
\bezier{50}(-250,150)(-200,175)(-250,200)
\bezier{50}(-250,200)(-300,225)(-250,250)
\bezier{50}(-250,250)(-200,275)(-250,300)
\bezier{50}(-250,300)(-300,325)(-250,350)
\bezier{50}(-250,350)(-200,375)(-250,400)
\put(-200,-400){\makebox(400,100){\Huge{Fig.7}}}
\end{picture}
\end{minipage}}
\end{picture}
\end{document}